# Broadband X-ray Burst Spectroscopy of the FRB-Emitting Galactic Magnetar

G. Younes[1,2,†], M. G. Baring[3,†], C. Kouveliotou[1,2,†], Z. Arzoumanian[4], T. Enoto[5], J. Doty[6], K. C. Gendreau[4], E. Göğüş[7], S. Guillot[8], T. Güver[9], A. K. Harding[10], W. C. G. Ho[11], A. J. van der Horst[1,2], C.-P. Hu[5], G. K. Jaisawal[12], Y. Kaneko[7], B. J. LaMarr[13], L. Lin[14], W. Majid[15], T. Okajima[4], J. Pope[4], P. S. Ray[16], O. J. Roberts[17], M. Saylor[4], J. F. Steiner[18], Z. Wadiasingh[4]

[1]Department of Physics, The George Washington University, Washington, DC 20052, USA

[2]Astronomy, Physics and Statistics Institute of Sciences (APSIS), The George Washington University, Washington, DC 20052, USA

[3]Department of Physics and Astronomy, Rice University, MS-108, P.O. Box 1892, Houston, TX 77251, USA

[4]Astrophysics Science Division, NASA GSFC, 8800 Greenbelt Rd., Greenbelt, MD 20771, USA

[5]Extreme Natural Phenomena RIKEN Hakubi Research Team, RIKEN Cluster for Pioneering Research, 2-1 Hirosawa, Wako, Saitama 351-0198, Japan

[6]Noqsi Aerospace, ltd 15 Blanchard Ave Billerica, MA 01821, USA

[7]Sabancı University, Orhanlı-Tuzla, İstanbul 34956, Turkey

[8]IRAP, CNRS, UPS, CNES, 9 avenue du Colonel Roche, BP 44346, F-31028 Toulouse Cedex 4, France

[9]Istanbul University, Science Faculty, Department of Astronomy and Space Sciences, Beyazıt, 34119, Istanbul, Turkey

[10]Theoretical Division, Los Alamos National Laboratory, Los Alamos, NM 58545

[11]Department of Physics and Astronomy, Haverford College, 370 Lancaster Avenue, Haverford, Pennsylvania 19041, USA

[12]National Space Institute, Technical University of Denmark, Elektrovej 327-328, DK-2800 Lyngby, Denmark

[13]MIT Kavli Institute for Astrophysics and Space Research, MIT, 70 Vassar Street, Cambridge, MA 02139, USA

[14]Department of Astronomy, Beijing Normal University, Beijing 100088, China

[15]Jet Propulsion Laboratory, California Institute of Technology, Pasadena, CA 91109, USA

[16]U.S. Naval Research Laboratory, Washington DC 20375, USA

[17]Universities Space Research Association, 320 Sparkman Drive, Huntsville, AL 35805, USA

[18]Harvard-Smithsonian Center for Astrophysics, 60 Garden St., Cambridge, MA, 02138, USA

[†]GY, MGB, and CK are co-corresponding authors of the manuscript. MGB and CK contributed equally and are listed alphabetically. Corresponding emails: gyounes@gwu.edu, baring@rice.edu, ckouveliotou@gwu.edu, respectively.

**Magnetars are young, magnetically-powered neutron stars possessing the strongest magnetic fields in the Universe. Fast Radio Bursts (FRBs) are extremely intense millisecond-long radio pulses of primarily extragalactic origin, and a leading attribution for their genesis focuses on magnetars. A hallmark signature of magnetars is their emission of bright, hard X-ray bursts of sub-second duration. On April 27$^{th}$ 2020, the Galactic magnetar SGR J1935+2154 emitted hundreds of X-ray bursts in a few hours. One of these temporally coincided with an FRB, the first detection of an FRB from the Milky Way. Here we present spectral and temporal analyses of 24 X-ray bursts emitted 13 hours prior to the FRB and seen simultaneously with NASA's NICER and Fermi/GBM missions in their combined energy range, 0.2 keV – 30 MeV. These broadband spectra permit direct comparison with the spectrum of the FRB-associated X-ray burst (FRB-X). We demonstrate that all 24 NICER/GBM bursts are very similar temporally, albeit strikingly different spectrally, from FRB-X. The singularity of the FRB-X burst is perhaps indicative of an uncommon locale for its origin. We suggest that this event originated in quasi-polar open or closed magnetic field lines that extend to high altitudes.**

SGR J1935+2154 was discovered in 2014, when it emitted a few short, hard X-ray, magnetar-like bursts. Follow-up X-ray observations revealed the source spin period (P=3.24 s) and period derivative (dP/dt=1.43x10$^{-11}$ s/s). Attributing this spin evolution to magnetic dipole torques on the rotation of the neutron star, a standard practice for pulsars and magnetars[1,2], it implies a very large surface dipole magnetic field, B~2.2x10$^{14}$ G, and a spin down age, τ=3.6 kyr, thus confirming the magnetar nature of this source[3]. The source became active again[4] in May 2015, May and June 2016, and December 2019. The source activity steadily increased with time, emitting larger numbers of bursts, brighter on average than the ones detected during the preceding activation[5]. On April 27$^{th}$ 2020, SGR J1935+2154 entered yet another active period, the most prolific so far. It comprised a long-lasting burst storm, with at least a few hundred bursts observed within a few hours[6-8]. The 1-25 keV persistent emission of the source was significantly enhanced subsequent to this storm over a period lasting several weeks[7,9].

We observed SGR J1935+2154 with the NICER X-ray Timing Instrument[10] (0.2-12 keV) onboard the International Space Station on April 28, from 00:40:58 UTC until 00:59:36 UTC (~19 minutes), covering just the tail end of the storm. This NICER observation revealed over 200 bursts[7] emitted by SGR J1935+2154, which was also visible to the Fermi Gamma-ray Burst Monitor (GBM; 8 keV – 30 MeV). Thirteen hours after the NICER observation, concurrent with a magnetar X-ray burst[11-13], a Fast Radio Burst (FRB) was detected with the CHIME[14] and STARE2[15] radio telescopes, though no persistent pulsed radio emission was observed in subsequent observations by FAST[16]. The FRB-associated X-ray burst (FRB-X) was detected by the INTEGRAL[11], KONUS-WIND[12], and HXMT[13] missions; NICER and GBM were not observing the source during that time. Although the FRB was at the faint end of the luminosity range typically encountered for extragalactic radio bursts[17,18], it firmly establishes the FRB-magnetar connection that has emerged as a popular paradigm[19].

As inferred from all the instruments that observed the FRB-X, its spectral characteristics were unusual, showing a harder, non-thermal profile, compared to bursts from previous activations[12] of SGR J1935+2154. To determine whether this was indeed spectrally distinct from the bursts observed around the time of the FRB, and to better characterize the nature of this spectral difference, we selected a subset of 24 bursts simultaneously detected with NICER and GBM, which afforded a broad-band energy coverage, sampling the full curvature of their X-ray spectra. Due to the high background of GBM, these were also the brightest among the 200 detected with NICER. As we did not observe the FRB-X, we compare our results to those inferred from the HXMT satellite, since it also covers the full broad-band energy span (1-300 keV) of the magnetar burst spectra.

We used the NICER data for a temporal analysis of the 24 X-ray bursts, as it offers a very low background compared to GBM, and hence captures the full length of each burst. The T90 duration[20] (interval during which 90% of the burst fluence is detected) of these bursts ranged from 230 ms to about 2 seconds, with a mean of 620 ms (Extended data Figure 1). The burst light curves display a variety of shapes, with some exhibiting a slow rise and decay bracketing a spiky top. Regarding its duration and temporal profile, the FRB-X does not stand out, compared to the 24 bursts.

We next performed time-integrated spectral analysis of the bursts using the combined data of NICER and GBM. We fit the broad-band spectra of all 24 bursts with, either a fully thermal model consisting of two blackbody (2BB) components, or a non-thermal model consisting of a power-law (PL) with a high-energy exponential cutoff (CPL), both modified by absorption from the interstellar medium along the line of sight to the source. We find that most of our bursts are adequately fit with both models, but also note that the 2BB model comprises one extra free parameter compared to CPL. Overall the CPL spectral model fits 23 of the 24 bursts consistently well, being superior to the 2BB model for 3 of these bursts (see methods). However, for the brightest burst, the statistically preferred fit was a non-purely thermal model (BB+CPL).

We present in Figure 1 the broadband spectrum of a burst with similar time-averaged flux to the FRB-X; its spectrum is typical for 23 of the 24 bursts. For comparison, we overlay in dashed lines the best-fit CPL model to a NICER+GBM simulated spectrum based on the HXMT FRB-X. The two spectra differ markedly, with the latter exhibiting a much higher cutoff energy and a significantly steeper PL component. This difference is intrinsic to the bursts: our simulations confirm that we would have easily detected and recovered to the few percent level the spectral parameters of a burst similar to the FRB-X (Extended data Figure 2).

Figure 2 demonstrates this difference with the distribution of the photon indices of the CPL model (left panel). Assuming that the HXMT burst is drawn from our sample of 24 spectroscopically similar bursts, for the photon index we measure a joint cumulative distribution function between the Kernel density function of our population of bursts and the probability density function (PDF) of the HXMT burst of about $1.42 \times 10^{-4}$. A similar analysis for the high-energy cutoff, $E_{cut}$, (Figure 2, right panel) implies that the probability of a burst with $E_{cut} = 84$ keV to be drawn from our $E_{cut}$ population is negligible ($1.0 \times 10^{-16}$). Finally, for all bursts we find a strong correlation between their cutoff energy and flux (Figure 3), with brighter bursts exhibiting higher energy cutoffs[12]. In both figures it is fairly obvious that the HXMT burst is an extreme outlier relative to our NICER+GBM sample.

The uniqueness of the FRB-X compared to the rest of the SGR J1935+2154 bursts extends beyond this recent activation. The GBM bursts from previous activations had average cutoff energies (CPL fits) of 16 keV, with a standard deviation of 3 keV, and photon indices of 0.1, with a standard deviation of 0.5[5]. Note that both parameters suffer from systematic uncertainties when measured in the GBM 8-200 keV energy range only, as is the case for the previous activations. A very recent analysis[8] of 148 bursts associated with the 2020 active epoch of SGR J1935+2154 presents a similar message of spectral softness. These and all earlier events are, therefore, consistent with our sample of 24 bursts within 1σ uncertainty, and further highlight the spectral dissimilitude of the FRB-X.

How can this special FRB-X exhibit such drastically distinct spectral properties? The answer may lie in its locale. The 0.2-2 sec durations of the 24 bursts constitute many regional transit timescales $R/c < \sim 0.3$ ms (for an emission region size $R \sim 10^6$-$10^7$ cm), implying that closed field lines are needed[21,22] to trap the plasma. If these field lines possess a fairly restricted range of altitudes R, the high opacity plasma[22] powering the emission will likely possess only a modest range of effective temperatures[23,24]. Smaller, hotter regions reside nearer the field line footpoints on the neutron star surface, i.e., for $R \sim R_{NS} \sim 10^6$ cm, (where $R_{NS}$ is the neutron star radius); altitudinal temperature gradients broaden the spectrum somewhat[24]. For a representative burst X-ray luminosity of $L_\gamma \sim 10^{40}$ erg/sec, simple invocation of the Stefan-Boltzmann law $L_\gamma = \sigma T^4 R^2$ yields a temperature of $T \sim 10^8$ K for $R \sim 10^6$ cm, commensurate with a value of $E_{cut} \sim 10$-15 keV, while at $R \sim 10^7$ cm altitudes, $T \sim 3 \times 10^7$ K. Accordingly, an altitude range spanning a decade yields a spectral extent compatible with the NICER-Fermi/GBM observations. This geometry for the emission regions of the 24 bursts could be provided by quasi-equatorial dipolar magnetic flux tubes, quadrupolar field morphologies or even toroidal structures associated with field line twists[25,26], all of which would present large emission solid angles, $\Omega \sim 2\pi$, to an observer at infinity.

The high $E_{cut} \sim 84$ keV and spectral breadth for the FRB-X suggest a temperature range spanning a decade or so, and therefore a much larger range of altitudes R, perhaps a factor 100 or more. This conceivably signals a locale for the activated field lines (open or closed) over the magnetic pole (see Methods). Magnetic trapping would then have less of an altitudinal "iso-thermalization" imprint and more of a collimating one with $\Omega << 4\pi$. The super-Eddington luminosity[21,22] would drive a mildly-relativistic flow upward from the stellar surface[22]. As this wind cools adiabatically before becoming transparent to electron scattering, the X-ray spectrum would soften, with the time-integrated convolution generating similar $\sigma T^4 R^2 \Omega / 4\pi$ effective luminosities over a broader range of photon energies. The high $E_{cut} \sim 84$ keV suggests $T \sim E_{cut}/k \sim 10^9$ K at the $R \sim 10^6$ cm base, implying $\Omega/4\pi \sim 10^{-4}$-$10^{-3}$, i.e., an opening angle of $\sim 1$-$3°$. At higher altitudes, the plasma would be unencumbered by magnetic Thomson scattering opacity and free to engage in coherent radio emission mechanisms. When convolved with the rotation of SGR J1935+2154, an FRB emission zone collimated by field lines to within a "cone" subtending around 3° can naturally produce (see Methods) radio variations on timescales of ~30ms.

Finally, the picture of the FRB and its X-ray counterpart FRB-X emanating from quasi-polar locales is supported by evidence[7] that their arrival times are coincident with the peak of the soft X-ray pulse profile where the surface thermal emission dominates. Such magnetar pulsation peaks are widely presumed[27-29] to correspond to phases where a polar hot spot on the stellar surface fleetingly faces towards a distant observer.

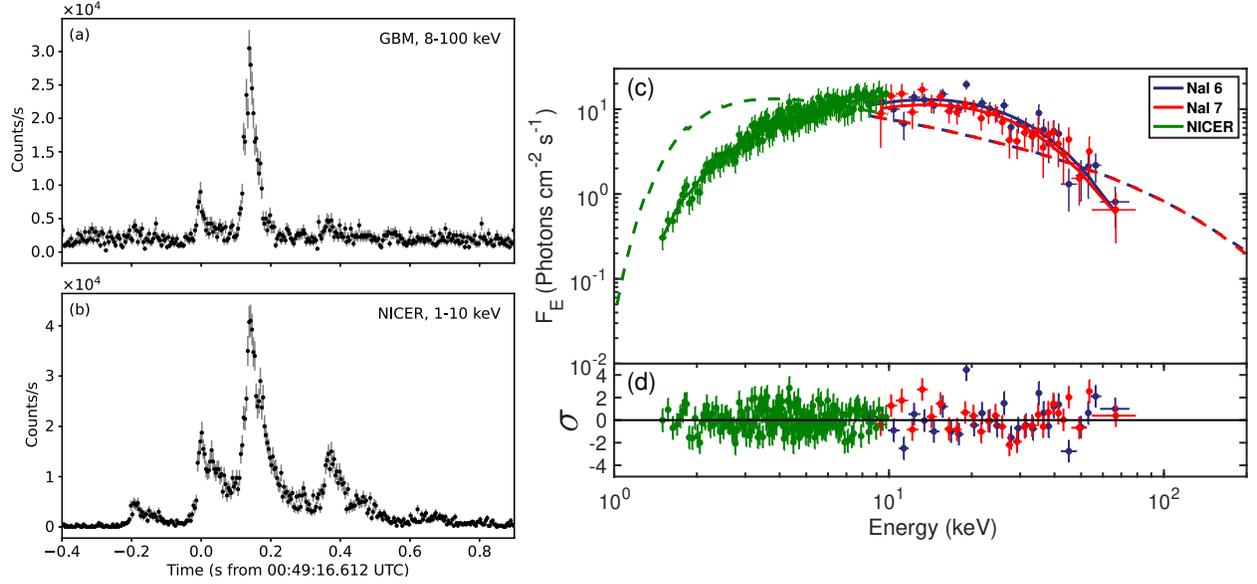

**Figure 1.** Example light curve and spectrum of one of the 24 bursts simultaneously observed with Fermi/GBM and NICER. Panel (a): Fermi/GBM light curve in the 8-100 keV range. Panel (b): NICER light curve in the 1-10 keV range. In both panels, the light curves are shown at the 4 ms resolution. The X-axis is time in seconds from a fiducial burst start time. The Y-axis is the number of counts per second. The black dots are the data points, and the corresponding dark gray lines are the 1 σ uncertainties. Panel (c): NICER+GBM spectrum of this example burst in photon flux space, $F_E$. The dots and corresponding vertical lines represent the spectral data and their corresponding 1σ uncertainty. The data is binned for clarity and color-coded by instrument (NaI 6, NaI 7 are the two GBM detectors used for this burst). The solid curves define the best-fit cutoff PL model. The dashed lines constitute the best-fit cutoff PL model to a simulated spectrum based on the spectral properties of the FRB-X as seen with HXMT[11]. This fit had spectral parameters of an index $\Gamma = 1.5 \pm 0.03$ and $E_{cut} = 84 \pm 9$ keV. Panel (d): residuals of the best-fit model to our NICER+GBM spectrum in standard deviation units (σ).

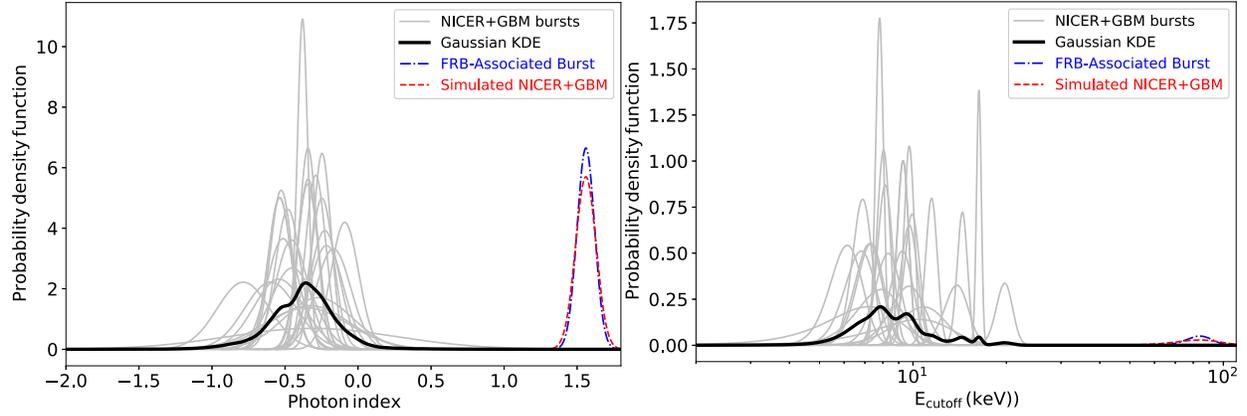

**Figure 2.** Spectral parameter distributions for all 24 bursts in our sample, and comparison to the FRB-X. Grey-solid lines represent the probability density function (PDF) of the cutoff PL index (left panel) and high-energy cutoff $E_{\rm cut}$ (right panel) for our sample of 24 bursts. In both panels, the black-solid lines are the PDF of a Gaussian kernel for the corresponding 24 PDFs. The blue dot-dashed lines are the PDFs of the index (left) and the high-energy cutoff (right) as measured with HXMT in the FRB-X. The red dashed lines are the PDFs of the index and cutoff energy of NICER+GBM simulated spectra based on the spectral parameters of the FRB-X (see Methods). The probability of the FRB-X to have an index drawn from our population of bursts is $1.4 \times 10^{-4}$, while the probability of $E_{\rm cut}$ to be drawn from our sample is $1.0 \times 10^{-16}$, highlighting the unique properties of the FRB-X compared to the rest of the burst population.

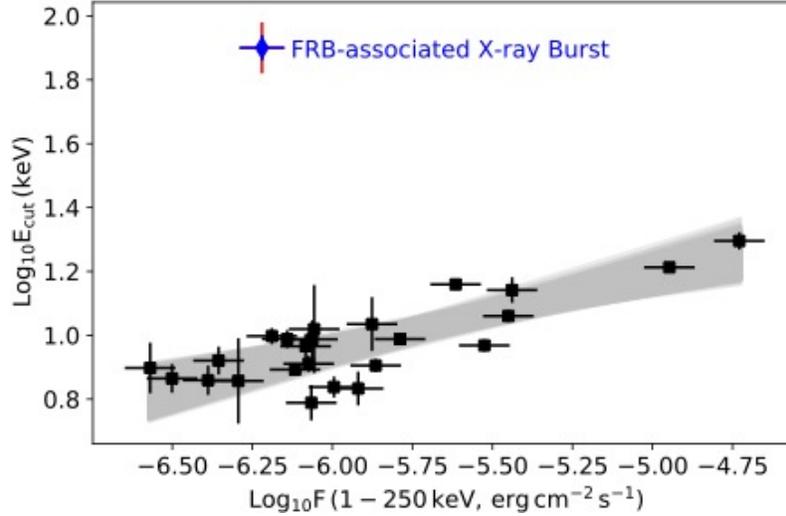

**Figure 3.** Cutoff energy, $E_{cut}$, versus flux in the 1-250 keV range for the 24 bursts in our sample. The black squares are the data points, while the black lines represent their uncertainty. A 20% systematic uncertainty was added to all flux values (see methods). The grey-shaded area is the 3σ best fit linear model to 10000 simulated sets of data points drawn from a bivariate Gaussian distribution with mean and standard deviation as measured in the actual data points. A positive correlation is clearly seen in our sample. The flux and $E_{cut}$ values of the FRB-X as measured with HXMT is shown as a blue-diamond. The red error bar is derived through GBM+NICER simulations based on the best-fit HXMT model of the FRB-X (see Methods). While possessing a typical flux, the $E_{cut}$ of the FRB-X is >15σ away from this correlation. We do not detect any other statistically significant correlation between any other pairs of spectral parameters in our sample.

**Data Availability Statement.** NICER raw data and cleaned, level-2 data files were generated at the Goddard Space Flight Center large-scale facility. These data files, with observation ID 3020560101, can be found at https://heasarc.gsfc.nasa.gov/FTP/nicer/data/obs/2020_04/3020560101. Fermi/GBM data files were generated at the Marshall Space Flight Center large-scale facility, and can be found at https://heasarc.gsfc.nasa.gov/FTP/fermi/data/gbm/daily/2020/04/28/current. Level 3 data supporting the findings of this study are available from the corresponding authors upon request.

**Code Availability Statement.** Reduction and analysis of the data were conducted using publicly available codes provided by the High Energy Astrophysics Science Archive Research Center (HEASARC), which is a service of the Astrophysics Science Division at NASA/GSFC and the High Energy Astrophysics Division of the Smithsonian Astrophysical Observatory. For NICER, NICERDAS version 7a, part of HEASOFT 6.27.2 (https://heasarc.gsfc.nasa.gov/docs/software/lheasoft) was used. Fermi tools version 1.0.3 and GSPEC version 0.9.1 were used for the analysis of Fermi/GBM data (https://fermi.gsfc.nasa.gov/ssc/data/analysis/gbm/). Spectral analysis was conducted using Xspec version 12.11.0 (https://heasarc.gsfc.nasa.gov/docs/xanadu/xspec/). Custom codes used to create plots presented in this manuscript are available from the corresponding authors upon request. These used python libraries NumPy[42], SciPy[43], and Matplotlib[44].

**Competing Interests.** The authors declare that they have no competing financial interests.

**Acknowledgments.** A portion of this work was supported by NASA through the NICER mission and the Astrophysics Explorers Program. This research has made use of data and software provided by the High Energy Astrophysics Science Archive Research Center (HEASARC), which is a service of the Astrophysics Science Division at NASA/GSFC and the High Energy Astrophysics Division of the Smithsonian Astrophysical Observatory. GY acknowledges support from NASA under NICER Guest Observer cycle-1 program 2098, grant number 80NSSC19K1452. CK acknowledges support from NASA under Fermi Guest Observer cycle-10 grant 80NSSC17K0761. MGB acknowledges the generous support of the NSF through grant AST-1813649. TE is supported by JSPS/MEXT KAKENHI grant numbers 18H01246, 18H04584, and the RIKEN Hakubi project. SG acknowledges the support of the CNES. WCGH acknowledges support from NASA through grant 80NSSC20K0278 NICER research at NRL is supported by NASA.

**Author contributions.** GY performed the data reduction and analysis and contributed to the writing of the manuscript. MGB contributed to the theoretical interpretation of the results and to the writing of the manuscript. CK contributed to the interpretation of the results and to the writing of the manuscript. ZA is the NICER project scientist. He contributed to the scheduling of the NICER observation, discussion on specific data analysis related to deadtime, discussion and editing of the paper. TE is the chair of the NICER magnetar and magnetosphere science group and responsible for observation planning of magnetars with NICER. JD designed the NICER electronics and contributed to the discussion related to data analysis and deadtime in the NICER data. KCG is the NICER PI; he approved the DDT observation that led to the detection of the burst storm with NICER. Authors EG, SG, TG, AKH, WCGH, AJH, CPH, GKJ, YK, LL, PSR, OJR, WM, and ZW contributed to the discussion and editing of the manuscript. BJL and JFS contributed to the discussion related to deadtime in NICER. TO is the NICER optics lead. JP and MS are part of NICER operations team who worked to get NICER on SGR J1935+2154 as quickly as possible.

## Methods

**NICER observations and data processing.** NICER[30] is a non-imaging instrument onboard the International Space Station, with a restricted field of view covering about 30 arcminutes$^2$. It consists of 56 coaligned X-ray concentrating optics, each with an associated Focal Plane Module (FPM) detector, 52 of which are currently operating. NICER is sensitive to photon energies in the range of 0.2-12 keV, and currently provides the largest collecting area in this energy band peaking at ~1900 cm$^2$ at 1.5 keV. The FPMs are split into groups of 8, which are simultaneously controlled by a set of electronics called a Measurement Power Unit (MPU). Each MPU operates independently of the others. For data reduction and processing, we use NICER Data Analysis Software (NICERDAS) version 7, as part of HEASoft 6.27.2, and Xselect version 2.4.

NICER observed SGR J1935+2154 on April 28$^{th}$ 2020 with several uninterrupted snapshots. The first covers the period from 00:40:58 UTC until 00:59:36 UTC, or approximately 19 minutes. Over 200 bursts were observed with NICER during this period. Our main focus in this Article is the analysis of the bursts that are simultaneously observed with GBM. Given the high background in GBM, the subset of 24 bursts employed in this analysis were the brightest bursts observed with NICER, and for some of these instrumental deadtime is non-negligible. Deadtime in NICER starts becoming significant for sources with count rates larger than 20000 counts s$^{-1}$, hence for integrations of tens of milliseconds, i.e., during the peak of the bursts, deadtime correction is required. We applied our deadtime correction by following the method described in Wilson-Hodge et al.[31]; here we give a summary of the steps. We start our analysis with the unfiltered event files for each MPU separately, applying standard filtering criteria to create corresponding filtered event files. We account for two types of deadtime, (1) the time during which each FPM of each MPU is "dead" while processing an event and (2) data packets lost due to saturation in each MPU slice. The first type of deadtime is recorded as a column in the event files, and we used the unfiltered event files to track it during the burst times (given that all events, not only the good ones, contribute to this type of deadtime). The second type of deadtime is recorded in the Good Time Intervals (GTIs) of the filtered event files and packet number in the housekeeping files for each MPU. This loss of events is apparent in the tails of the two brightest bursts (bursts 3 and 8 in Table 1); however, it does not affect any of the other 22 bursts we analyze here.

For our spectral analysis, a deadtime-corrected exposure for each burst is derived after correcting for the fraction of exposure that is lost due to the two types of deadtime mentioned above. We find that deadtime is most significant for the two brightest bursts with the lost GTIs, and we estimated a deadtime fraction of about 30% and 20%, respectively. For the remaining 22 bursts, the deadtime fraction ranged from 10 to about 2 percent.

Given NICER's comparatively small field of view, the background for the bursts' spectra was assumed to be the underlying burst-free persistent emission, most probably originating on the stellar surface. This component varies throughout the observation, and hence, was measured in segments of 100 seconds, constituting around 31 stellar rotation periods. This background constituted less than 1% of the fluxes for all 24 bursts. Finally, we use the NICER response files provided in the HEASoft calibration database, version 20200722.

**Fermi-GBM observations and data reduction.** The Gamma-ray Burst Monitor[32] (GBM) onboard the Fermi Gamma-ray Space Telescope consists of 12 Sodium-Iodide (NaI) detectors sensitive to photons in the energy range 8-1000 keV and 2 Bismuth Germanate (BGO) detectors sensitive to photons in the 0.2-40 MeV range. The detectors are spread over a cubic configuration which covers the full Earth un-occulted sky. SGR J1935+2154 was in the field of view of GBM during the entire length of the first NICER snapshot. Few detectors had good viewing angles towards the source (<50 degrees) without any blockage from the spacecraft itself. We use these detectors for our spectral and temporal analyses. GBM automatically triggered on only one occasion during the NICER observation, hence, we relied on the continuous time-tagged events (CTTE), with a time resolution of 2 microseconds, to search for other bursts that were detected with NICER in the same time span. We extracted burst and background spectra using the GBM Data Tools version 1.0.3 and created response files for each burst using the GBM Response Generator version 2.0, which uses GBM calibration files version 10.

**Burst search.** We performed a burst search in both NICER and GBM in a similar manner. The search consisted of estimating the Poisson probability of a time-bin (with a certain resolution, *tbin*) to be a random fluctuation around an average mean within a certain time-interval (*dt*). Any *tbin* with counts that show >5σ deviation from the mean is saved as a possible burst[33]. The procedure is repeated after excluding all bins that were flagged as bursts, until no further bins are found to deviate sufficiently from the mean. We experimented with multiple time-intervals *dt*, namely between 20 and 200 seconds in steps of 20 seconds and found that they all gave consistent results. Our final results are for *dt*=100 seconds. We performed the search using multiple time resolutions (4 ms, 32 ms, 128 ms, and 512 ms) so that we do not miss any possible weak precursors or faint tails to the bursts. For NICER, we performed the search on all 52 FPMs combined. For GBM, we performed the search on each of the NaI detectors separately. In both NICER and GBM, two bursts were considered separate if the count rate between them remained at the background level for 0.5 seconds or longer. This corresponds to less than 15% of the magnetar spin rotation period. Using this method, we find over 200 bursts in NICER and 24 bursts in GBM. All GBM bursts were also found in NICER.

**Temporal analysis.** We measured the T90 duration[20] for each burst, i.e., the interval of time during which 5% to 95% of the total burst fluence is accumulated. Given the very low count background of NICER compared to GBM, we relied on the former data to estimate T90s, since the latter would underestimate the T90. We performed this analysis in count space. We built light curves at 4-millisecond resolution and corrected the number of counts in each 4 millisecond bin for the loss of exposure due to deadtime. We estimated the background in intervals of 0.5 to a few seconds just before and after the start and end times of the bursts, respectively. We created a background-corrected cumulative counts plot and assumed that the burst T100 (or 100 percent of the burst fluence) resides 3σ above and below the background before the start and after the end of the burst, respectively. Then, we estimated the T90 from this background-corrected interval. The distribution of T90s for the 24 NICER+GBM bursts is shown in Figure 1 (supplement). The distribution is broad with a mean of about 620 ms. Hence, the T90 duration of the FRB-X as measured with HXMT[13], which is about 530 ms, is well within the population of bursts as observed with NICER. Note that the instruments on board HXMT are low background instruments and hence more appropriately compared to NICER rather than GBM. We also note that the bursts' temporal shapes as observed with NICER are considerably varied, with a few closely resembling the FRB-X[13]: a slow rise and a slow decay, separated by a spiky structure.

**Spectral analysis.** We performed our spectral analysis using the X-ray Spectral Fitting Package Xspec version 12.11.0k[34]. For each burst, we simultaneously fit the NICER spectrum and the spectra from all GBM detectors that satisfied the criteria as described above. For each burst, we fit for the time interval T90 as measured with GBM to maximize the signal to noise ratio at high energies. This corresponds to an average of ~70% of the full length of the NICER bursts. We verified that performing our spectral analysis using the NICER T90 does not alter any of our conclusions. For all spectral models described below, we add an absorption component due to the interstellar medium between Earth and SGR J1935+2154. For this purpose, we used the *tbabs* model in Xspec. We assumed the abundances of Wilms et al.[35] and the photo-electric cross-sections of Verner et al[36]. Moreover, we add a multiplicative constant to all the models to take into account any calibration uncertainties between all the instruments. We find this constant normalization to be at most 10% between the GBM detectors. As for the difference between NICER and GBM, we find this calibration uncertainty to be between 10 and 60%, with the highest deviations (also with the largest uncertainties) corresponding to the weakest bursts. The average of this calibration uncertainty among our population of 24 bursts is 25±20%.

We use the *pgstat* statistics in Xspec to estimate the best-fit model parameters and their associated uncertainties. This statistic is usually used for Poisson distributed data with Gaussian distributed background: the case of our spectra. To test the goodness-of-fit for each model, we relied on the Anderson-Darling[37] (AD) test statistic, which compares the empirical distribution functions of the data and model (details on these statistics can be found in https://heasarc.gsfc.nasa.gov/xanadu/xspec/manual/XSappendixStatistics.html). We utilize the goodness command in Xspec to simulate 1000 spectra based on a given model and compare their AD test statistic to that of the data. If the data are drawn from this model, then around 50% or less of the simulated spectra should have test statistic less than that of the data.

We first fit the spectra with a simple model consisting of either a blackbody (BB) or a power-law (PL). Individually, these two simple models failed to give a statistically good fit to any of our 24 bursts. We then fit the data with the two principal models that are usually invoked to explain the spectral curvature of magnetar short bursts. These are the two blackbodies (2BB) and a cutoff PL (CPL), the latter possessing one less free parameter. According to the simulations as described above, the CPL model gave consistently good fits to 23 bursts, barring the brightest one in our sample. On the other hand, the 2BB model resulted in either similar goodness-of-fit results compared to the CPL model or slightly worse (e.g., bursts 1, 8, and 14). The brightest burst (flagged in Table 1 supplementary material with an asterisk) cannot be adequately explained with either of the above two models, although the CPL results in better statistics compared to the 2BB one. For that burst, we find that the combination of a BB+CPL model is required to give a good fit. We conclude that, given the smaller number of parameters for the CPL model compared to the 2BB and its moderately better performance across flux levels, the CPL model is adequate to describe 23 of the 24 bursts that we analyze in this Article. An extra BB component with temperature T=8.6±0.3 keV is required for the brightest one. The spectral parameters for all these bursts are summarized in Table 1. The fluxes are given in the 1-250 keV band for direct comparison with the flux of the FRB-X[13]. Therein, we only quote the results of the CPL parameters, although we give the results of our simulations to gauge the goodness of fits from both the CPL and the 2BB model for ease of comparison (last two columns).

To examine the detectability of the FRB-X and its spectral appearance with the combined NICER+GBM instruments, we performed a set of simulations using the *fakeit* command in Xspec as follows. Using the response matrices of the GBM detectors that had good viewing angles towards the source, and the background as estimated from the actual data around the time of the burst storm, we simulated two NaI spectra (same number as the detectors that had good viewing angle to the source throughout our burst storm coverage) using the best fit CPL model parameters as derived with HXMT ($\Gamma = 1.56$, $E_{cut} = 84$ keV, $N_H = 2.7 \times 10^{22}$ cm$^{-2}$, and $F = 6.0 \times 10^{-7}$ erg s$^{-1}$ cm$^{-2}$). We repeated the same procedure for NICER. We then fit the two simulated GBM spectra and the NICER simulated spectrum with a CPL (this best fit model is shown as dashed lines in Figure 1). The parameters of our best-fit model are $\Gamma = 1.55 \pm 0.03$, $E_{cut} = 86 \pm 9$ keV, and $N_H = (2.69 \pm 0.04) \times 10^{22}$ cm$^{-2}$, all of which are consistent with HXMT derived parameters at the $1\sigma$ level. The PDF of the index and the cutoff energy are shown as red dashed lines in Figure 2.

We then simulated 10000 GBM+NICER spectra as described above, but with spectral parameters drawn from Gaussian distributions with mean and standard deviation as measured with HXMT. We fit each of these spectra with a CPL. The distribution of the best-fit spectral parameters, shown in Extended Data Figure 2, follow a normal distribution. We find $\Gamma = 1.56 \pm 0.07$, $E_{cut} = 84 \pm 8$ keV, and $N_H = (2.7 \pm 0.1) \times 10^{22}$ cm$^{-2}$. These are again consistent with the HXMT derived fit parameters at the 1σ level, and obviously inconsistent with the spectral shape of the 24 NICER+GBM bursts we analyze in this work.

**Interpretation - The Locales of the FRB and its X-ray burst:** If the FRB-X arises in a quasi-equatorial zone, at magnetic colatitudes greater than around 20°, the lower altitudes that likely produce the emission seen[12] at energies at around 100-200 keV (above $E_{cut}$) must exceed around $5 \times 10^6$ cm, in order to not be attenuated by photon splitting.[38] Then in order to generate the broad spectrum down to below 10 keV, using the Stefan-Boltzmann law, the outer realms of the burst zone likely would be at altitudes above R~$5 \times 10^8$ cm, i.e., 500 stellar radii $R_{NS}$, corresponding to light crossing times R/c~16ms, possibly conflicting with the <10ms variability timescales[13] of the FRB-X. The destructive influence of high Thomson opacity on coherence of electron populations likely precludes the radio emission region from being co-spatial with that of the contemporaneous X-ray burst. Accordingly, for quasi-equatorial locales for the FRB and its associated X-ray burst, the FRB would then likely originate at altitudes of greater than R~$5 \times 10^8$ cm. This renders it difficult to produce the FRB spikes separated by 29ms that are coincident[13] with peaks in the X-ray light curve and, therefore, presumably causally connected.

This conflict can be ameliorated by assuming that the FRB-X emanates from approximately polar locales. Then, since X-ray attenuation by photon splitting is generally much less near the poles[38], the maximum altitude for X-ray emission can be lower, nominally around 100 $R_{NS}$, and likely even less if it emanates from a collimated relativistic wind. Thus, a quasi-polar/non-polar dichotomy for the FRB-X and "orphan" X-ray bursts emerges, and is consistent with the rarity of FRB-X. Uniformly distributing the activation locales on the surface for hundreds of SGR J1935+2154 X-ray bursts establishes an average angular separation of their flux tube footpoint centroids of around 4-5°. The polar colatitude of the last open field line for this magnetar is $\theta_c \sim (2\pi R_{NS}/Pc)^{1/2} \sim 0.46°$. Accordingly, if the FRB-X is generated proximate to the open field line zone, it is essentially unique[39] in the archival burst assemblage.

It is notable that the two peaks of the FRB are separated by 29 ms[14,15], corresponding to a stellar rotation through angle 3.3°. Such temporal morphology of the radio signal is unlikely to come from highly-curved field lines at high quasi-equatorial altitudes. Yet it is a natural outcome of a highly-collimated emission region within a slightly flared flux tube over the pole. The angular extent of this zone must exceed around $\Delta\theta_e \sim 3-4°$ for the two radio peaks to be observed. Given the polar field line flaring relation $R/R_{NS} \sim (\Delta\theta_e/\theta_c)^2$ for dipolar field morphology, this implies an FRB emission locale at more than ~50-100 stellar radii $R_{NS}$, high enough to enable transparency to Thomson scattering. Detecting the FRB then requires the observer to approximately sample the magnetic pole, tilted relative to the spin axis, once during the rotation period; this special observational perspective naturally accounts for the rarity of luminous FRBs with associated X-ray bursts. Thus, emerges a paradigm of a high-altitude, quasi-polar locale for this FRB that is similar to the perceived site[40,41] for persistent radio emission in normal pulsars. This determination for radio pulsars is underpinned by rapid variations in the direction of their polarization on the plane of the sky[40], hallmarks of a viewing perspective almost along magnetic field lines.

**Table 1. Burst durations and spectral parameters.** Time is from 2020 April 28, 00h. The burst with an asterisk is the one burst where a BB+CPL model is required to provide a statistically good fit to the data. A 2BB model cannot provide a good fit as is evident from the last column. The burst highlighted in bold-face is the one presented in Figure 1 of the main text. Numbers in parentheses represent the 1$\sigma$ uncertainty on the corresponding last digit.

| Burst # | TIME UTC | T90 (ms) | $N_H$ $10^{22}$ cm$^{-2}$ | $\Gamma$ | $E_{cut}$ keV | $F_{1-250keV}$ $10^{-7}$ erg s$^{-1}$ cm$^{-2}$ | Constant | pgstat/dof | Goodness | Goodness (2BB) |
|---|---|---|---|---|---|---|---|---|---|---|
| 1 | 41:32.143 | 408(4) | 3.1(1) | -0.45(3) | 14(1) | 38.0(8) | 1.2(1) | 906/938 | 38.4 | 90.2 |
| 2 | 43:25.184 | 776(7) | 3.9(2) | -0.34(6) | 9.7(4) | 16.5(4) | 1.1(1) | 316/303 | 58.6 | 56.3 |
| 3* | 44:08.212 | 445(5) | 2.8(1) | -0.25(8) | 20(1) | 186.7(4) | 0.82(4) | 280/318 | 16.9 | 100 |
| 4 | 44:09.286 | 276(10) | 4.0(3) | -0.5(1) | 11.5(5) | 35(1) | 1.2(1) | 203/188 | 50.4 | 39.8 |
| 5 | 45:31.099 | 352(9) | 3.0(5) | -0.5(2) | 10(1) | 8.7(7) | 1.4(3) | 465/584 | 61.5 | 43.8 |
| 6 | 46:00.035 | 1160(20) | 3.8(2) | -0.30(7) | 9.7(6) | 7.2(3) | 1.3(3) | 278/273 | 21.5 | 65.5 |
| 7 | 46:06.427 | 231(7) | 3.9(7) | -0.27(4) | 11(2) | 13(1) | 1.4(3) | 317/419 | 45.6 | 48.4 |
| 8 | 46:20.170 | 654(8) | 3.1(1) | -0.38(4) | 16.3(3) | 112(1) | 0.87(4) | 460/374 | 70.9 | 100 |
| 9 | 46:23.456 | 1190(20) | 3.8(3) | -0.2(1) | 7.3(6) | 3.0(1) | 0.8(1) | 774/792 | 85.3 | 86.6 |
| **10** | **46:43.088** | **672(8)** | **3.3(4)** | **-0.34(1)** | **9.9(5)** | **6.5(2)** | **1.2(2)** | **226/225** | **30.4** | **32.5** |
| 11 | 47:24.977 | 875(5) | 4.5(5) | -0.45(11) | 8(1) | 2.7(3) | 1.4(5) | 507/627 | 31.8 | 82.8 |
| 12 | 47:57.532 | 741(7) | 4.0(5) | -0.2(1) | 9(1) | 8.3(3) | 1.2(3) | 758/736 | 59.7 | 10.7 |
| 13 | 48:44.836 | 652(8) | 3.7(2) | -0.2(1) | 8(1) | 4.4(2) | 1.4(2) | 812/836 | 91.3 | 48.3 |
| 14 | 48:49.270 | 985(22) | 3.9(3) | -0.25(6) | 14.3 | 24.3(7) | 1.3(1) | 854/933 | 75.3 | 100 |
| 15 | 49:00.275 | 2090(40) | 4.2(2) | 0.0(1) | 7(1) | 4.1(2) | 0.8(2) | 814/841 | 57.3 | 51.3 |
| 16 | 49:06.474 | 517(10) | 4.2(2) | -0.3(2) | 7(2) | 5.1(7) | 1.4(3) | 357/501 | 3.2 | 37.3 |
| 17 | 49:16.610 | 877(11) | 3.8(2) | -0.52(8) | 8.1(4) | 14.7(4) | 1.3(1) | 848/873 | 12.2 | 84.3 |
| 18 | 49:22.393 | 304(8) | 3.1(4) | -0.8(2) | 6(1) | 8.6(6) | 1.2(3) | 532/627 | 31.1 | 78.2 |
| 19 | 49:27.323 | 401(6) | 3.3(8) | -0.4(2) | 10(3) | 8.8(2) | 1.6(5) | 260/349 | 40.2 | 24.5 |
| 20 | 49:46.678 | 290(10) | 4.3(5) | -0.5(1) | 6.9(5) | 10.1(5) | 1.3(2) | 737/809 | 56.6 | 31.4 |
| 21 | 50:01.031 | 817(7) | 3.6(2) | -0.3(1) | 8.2(4) | 8.4(3) | 1.4(3) | 810/876 | 64.2 | 62.9 |
| 22 | 51:35.913 | 522(8) | 4.0(5) | -0.6(2) | 7(1) | 12.0(8) | 1.3(2) | 548/605 | 19.1 | 44.5 |
| 23 | 51:55.453 | 763(5) | 4.5(6) | -0.3(4) | 7.8(2) | 7.6(7) | 1.2(4) | 488/567 | 17.0 | 29.7 |
| 24 | 54:57.475 | 315(8) | 3.6(2) | -0.54(8) | 9.3(4) | 30(1) | 1.2(2) | 816/859 | 88.5 | 40.4 |

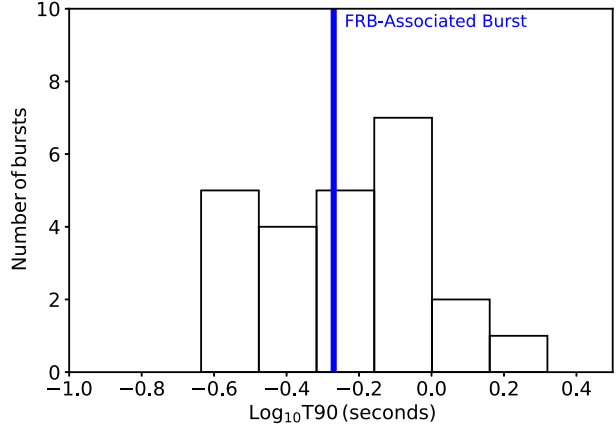

**Extended Data Figure 1.** T90 distribution of the 24 bursts in our sample. The blue bar represents the T90 of the FRB-X as measured with HXMT.[13]

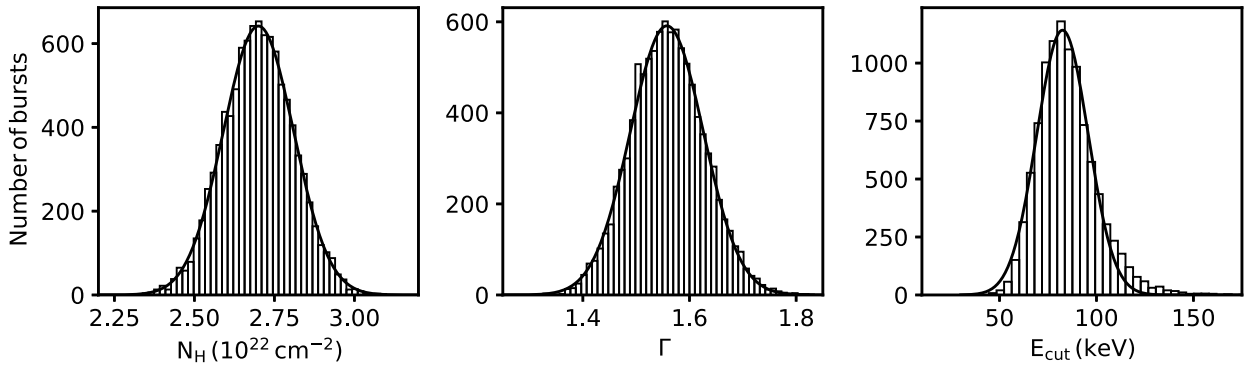

**Extended Data Figure 2.** Distributions of the spectral parameters of a CPL model that best fit 10000 simulated NICER+GBM spectra. The simulated spectra are drawn from the best fit CPL model to the HXMT FRB-X[13].

18